\documentclass[aps,preprint]{revtex4}%
\usepackage{amsfonts}
\usepackage{amsmath}
\usepackage{amssymb}
\usepackage{graphicx}%
\providecommand{\U}[1]{\protect\rule{.1in}{.1in}}
\begin{document}
\title{Statistical link between Bell nonlocality and uncertainty relations }
\author{Li-Yi Hsu \\
Department of Physics, Chung Yuan Christian University, Chungli 32081, Taiwan \\
and \\
Physics Division, National Center for Theoretical Sciences, Taipei 106319, Taiwan}

\begin{abstract}
As the distinguished features of quantum theory, Bell nonlocality concerns the
upper bound of the correlation strength of local observables on different
systems, and the uncertainty relations set the lower bound on the sum or
product of their variance on a quantum particle. Here, using the
Aharonov-Vaidman identity, we show the inextricable connection of these two
quantum properties by setting the statistical link between the degree of Bell
nonlocality and the joint uncertainty. Specifically, the upper bounds of Bell
inequalities depend on the joint measurement uncertainty. Creating a higher
degree of Bell nonlocality from an entangled state requires more measurement
uncertainty of local observables.

\end{abstract}
\volumeyear{year}
\volumenumber{number}
\issuenumber{number}
\eid{identifier}
\date[Date text]{date}
\received[Received text]{date}

\revised[Revised text]{date}

\accepted[Accepted text]{date}

\published[Published text]{date}

\startpage{1}
\endpage{2}
\maketitle

Two of the most distinguished features of quantum theory from classical
physics are uncertainty relations and nonlocality. Originally, Heisenberg
observed the inherent fact that position and momentum cannot be predicted
simultaneously with certainty, and the product of their variances is
approximately equal to the Planck constant \cite{Heisenberg}. Kennard
\cite{Kennard}, Robertson \cite{Robertson}, and Schr\"{o}inger \cite{Schro,
Schro2} further established the well-known uncertainty relations that define
the lower bounds of the product of variances of two non-commuting observables.
Over time, traditional uncertainty relations are formulated in terms of the
variance of measurement results \cite{M1,M2,M3,M4,M5,M6,M7}. With the rise of
quantum information science, variant entropies for depicting uncertainty in
information theory are widely introduced. Deutsch initially introduced the
entropic uncertainty relation in quantum measurements \cite{e1}, which was
improved by Maassen and Uffink \cite{e2}. Later, uncertainty relations with
the memory effect were proposed \cite{e3}. Different uncertainty relations
with various entropic functions for tasks in quantum cryptography, quantum
communication, and quantum thermodynamics \cite{e4, e5}. Entropic uncertainty
relations impose the lower bound of the extracted information that depends on
the incompatibility of the observables.

On the other hand, quantum nonlocality includes Bell nonlocality, quantum
steering, and quantum entanglement. Bell nonlocality is the strongest and most
well-known sort among them. In Bell tests, a source initially emits a state of
two or more particles that are received by spatially separated observers. In
each round of the measuring process, each observer performs a local
measurement with a random measurement setting as the input and then obtains
the measurement outcome as the output. Given the emitted quantum entangled
states, the input-output correlation cannot be reproduced using any local
hidden variable. According to the operational criteria of Bell nonlocality in
the Bell tests, Bell inequalities are always satisfied by any local hidden
model, while Bell nonlocality brings their violations. Uncertainty relations
and Bell nonlocality both have been established as important tools for a wide
range of applications in quantum technology. Both are widely used in quantum
cryptography \cite{qc1,qc2}. In addition, uncertainty relations are also used
in quantum metrology and quantum speed limit \cite{u1,u2,u3}, and Bell
nonlocality has led to applications in communication complexity reductions
\cite{cc}, private random number generation \cite{prng}, and self-testing
protocols \cite{self}.

Recently, the connection between these two central concepts has been
investigated. For example, to do this, the Bell test of the
Clauser-Horne-Shimony-Holt (CHSH) inequality is conceived as a nonlocal game,
and the fine-grained uncertainty relations as entropic ones represent the
maximal winning probability over all possible strategies \cite{OW}. Regarding
the winning information in a nonlocal game, it is shown that the strength of
the uncertainty determines the degree of nonlocality and there needs some
specific amount of uncertainty to achieve a certain nonlocality \cite{OW,ow3}.
Different entropies are used to evaluate accessible information to explore the
link between the steering effect and the uncertainty relations
\cite{ow1,ow2,oo}. An alternative way to find the connection between
nonlocality and measurement uncertainty from the no-signaling principle is
proposed \cite{pp}. Therein, the trade-off between the information gain in
measuring one observable and its disturbance on the remaining observables is discussed.

Here, we build the statistical link between these two features in standard
Bell tests. We introduce the proposed inequalities that involve statistical
characteristics of local observables rather than abstract entropic quantities
or monogamy relations. In quantum theory, there are upper bounds on Bell-type
inequalities and lower bounds on the sums or products of variances in the
uncertainty relations. To strike a balance, we propose the upper bound of Bell
nonlocality strength in terms of the variances and expectation values of the
local measurement outcomes. A few remarks on uncertainty relations are in
order. In the Bell test of a specific Bell inequality, each observer random
measures incompatible observables, and thus can also test local uncertainty
relations. Here, we focus on the conventional uncertainty relations based on
the sum of variances of the measurement results \cite{M3,M4,M5,M6,M7} since
those in terms of the product of the variances can be null and thus become
trivial even for incompatible observables. Given the observables $A$\ and $B$
on a single quantum system with the state $\left\vert \psi\right\rangle $, the
local uncertainty concerns the lower bound of the root of the local sum of the
variance square (RLSoVS) $(\bigtriangleup A)^{2}+(\bigtriangleup B)^{2}$,
which can be expressed in terms of quantities such as $\left\langle
\psi\left\vert \lbrack A\text{, }B]\right\vert \psi\right\rangle $,
$\left\langle \psi^{\perp}\left\vert A+iB\right\vert \psi\right\rangle $,
$\bigtriangleup(A+B)$ that, however, cannot be measured directly \cite{M3,
M4,M5}. On the other hand, regarding $\left\vert \psi\right\rangle $ as a
state vector of a composite quantum system, the observables $A$\ and $B$ are
used in the Bell test, and therefore each observer can obtain RLSoVS and
evaluate the correlation strength between the measurement observables. Hence,
the joint uncertainty as the product of RLSoVS and Bell nonlocality can
be\ linked using the statistics of local observables. Qualitatively speaking,
our study shows that a specific amount of joint uncertainty is required to
achieve a certain nonlocality, and hence a larger amount of joint uncertainty
allows for a higher degree of nonlocality.

The Aharonov-Vaidman identity is an important mathematical tool to establish
the statistical link between Bell nonlocality and uncertainty relations. We
review the Aharonov-Vaidman identity as follows. Given a Hermitian operator
$A$ on a Hilbert space $\mathcal{H}$. and a vector $\left\vert \psi
\right\rangle \in\mathcal{H}$, the Aharonov-Vaidman identity reads \cite{AV,
AV2}%

\begin{equation}
A\left\vert \psi\right\rangle =\left\langle A\right\rangle \left\vert
\psi\right\rangle +\triangle A\left\vert \psi_{A}^{\perp}\right\rangle .
\label{av}%
\end{equation}
Here $\left\langle A\right\rangle =\left\langle \psi\left\vert A\right\vert
\psi\right\rangle /\left\langle \psi|\psi\right\rangle $ and $\triangle
A=\sqrt{\left\langle A^{2}\right\rangle -\left\langle A\right\rangle ^{2}}$
represent the expectation value and the variance of the observable $A$,
respectively, and $\left\vert \psi_{A}^{\perp}\right\rangle $\ denote a state
vector orthogonal to $\left\vert \psi\right\rangle $. Eq. (\ref{av}) holds
only if $\left\langle \psi_{A}^{\perp}|\psi_{A}^{\perp}\right\rangle
=\left\langle \psi|\psi\right\rangle $, and hereafter all state vectors are
assumed to be normalized. According to the postulate of quantum theory, the
state vector determines the properties of a (composite) quantum system,
whereas the Aharonov-Vaidman identity depicts the statistics of the observable
$A$. The Aharonov-Vaidman identity helps to derive the Robertson uncertainty
relation and the stronger ones for all incompatible observables \cite{av 1,M4,
M5, Leifer}, explore the eccentric properties of weak values \cite{ad1}, and
determine the moments of a Hermitian operator \cite{ad2}. Regarding
$\left\vert \psi\right\rangle $ usually as a state vector of a single quantum
system, variant uncertainty relations between incompatible observables can be
derived using the Aharonov-Vaidman identity \cite{av 1,M4, M5, Leifer}.

In this work, we apply the Aharonov-Vaidman identity to composite quantum
systems. As an illustration, let the state vector $\left\vert \psi
\right\rangle $ be a two-qubit state, and the observables $A$\ and $B$ be the
observables on two different quantum systems. As a result, we have
\begin{equation}
\left\langle AB\right\rangle =\left\langle A\right\rangle \left\langle
B\right\rangle +\triangle A\triangle B\left\langle \psi_{A}^{\perp}|\psi
_{B}^{\perp}\right\rangle , \label{demo}%
\end{equation}
where $\left\langle \cdot\right\rangle $ denotes $\left\langle \psi\left\vert
\cdot\right\vert \psi\right\rangle $. The LHS of (\ref{demo}) consists of the
correlator of $A$ and $B$ that contributes to the correlation strength in the
Bell inequality, and the term $\left\langle AB\right\rangle -\left\langle
A\right\rangle \left\langle B\right\rangle $ reflects the correlation. On the
other hand, the RHS of (\ref{demo}) consists of the variances of the
observables that contribute to the uncertainty relations. In this way, we can
take advantage of (\ref{demo}) to establish the link between Bell nonlocality
and uncertainty relations. In the following discussion, $\left\vert
\psi\right\rangle $ can be a state vector of a multi-qubit or multi-qudit
system in Bell tests.

\textit{Two-qubit case \ }In the bipartite CHSH-Bell test, distant Alice and
Bob measure the local dichromatic observables denoted as $A_{x}%
=\overrightarrow{a}_{x}\cdot$ $\overrightarrow{\sigma}$ and $B_{y}%
=\overrightarrow{b}_{y}\cdot$ $\overrightarrow{\sigma}$, and the associated
Hermitian operators are denoted as $\widetilde{A}_{x}=A_{x}-\left\langle
A_{x}\right\rangle I$ and $\widetilde{B}_{y}=B_{y}-\left\langle B_{y}%
\right\rangle I$, respectively, where $I$ is the identity operator. The
CHSH-Bell operator is defined as $CHSH=%
%TCIMACRO{\dsum \nolimits_{x,y=0}^{1}}%
%BeginExpansion
{\displaystyle\sum\nolimits_{x,y=0}^{1}}
%EndExpansion
(-1)^{xy}A_{x}B_{y}$, and related CHSH-Bell operator is defined as
$\widetilde{CHSH}=%
%TCIMACRO{\dsum \nolimits_{x,y=0}^{1}}%
%BeginExpansion
{\displaystyle\sum\nolimits_{x,y=0}^{1}}
%EndExpansion
(-1)^{xy}\widetilde{A}_{x}\widetilde{B}_{y}$. Applying the Aharanov-Vaidman
identity, we have $\widetilde{A}_{x}\left\vert \psi\right\rangle =\triangle
A_{x}\left\vert \psi_{A_{x}}^{\perp}\right\rangle ,$ $\widetilde{B}%
_{y}\left\vert \psi\right\rangle =\triangle B_{y}\left\vert \psi_{B_{y}%
}^{\perp}\right\rangle $. This leads to the expression $\left\langle
\widetilde{A}_{x}\widetilde{B}_{y}\right\rangle =\triangle A_{x}\triangle
B_{y}\left\langle \psi_{A_{x}}^{\perp}|\psi_{B_{y}}^{\perp}\right\rangle $,
which is a real number \cite{Un}. We have
\begin{align}
&  \left\langle \widetilde{CHSH}\right\rangle \nonumber\\
&  =\left\langle CHSH\right\rangle -\left\langle CHSH\right\rangle
_{L}\nonumber\\
&  =\triangle A_{0}\left\langle \psi_{A_{0}}^{\perp}\right\vert (\triangle
B_{0}\left\vert \psi_{B_{0}}^{\perp}\right\rangle +\triangle B_{1}\left\vert
\psi_{B_{1}}^{\perp}\right\rangle )+\triangle A_{1}\left\langle \psi_{A_{1}%
}^{\perp}\right\vert (\triangle B_{0}\left\vert \psi_{B_{0}}^{\perp
}\right\rangle -B_{1}\left\vert \psi_{B_{1}}^{\perp}\right\rangle )\nonumber\\
&  \leq\triangle A_{0}\left\vert \triangle B_{0}\left\vert \psi_{B_{0}}%
^{\perp}\right\rangle +\triangle B_{1}\left\vert \psi_{B_{1}}^{\perp
}\right\rangle \right\vert +\triangle A_{1}\left\vert \triangle B_{0}%
\left\vert \psi_{B_{0}}^{\perp}\right\rangle -\triangle B_{1}\left\vert
\psi_{B_{1}}^{\perp}\right\rangle \right\vert \nonumber\\
&  \leq\sqrt{(\triangle A_{0})^{2}+(\triangle A_{1})^{2}}\sqrt{%
%TCIMACRO{\dsum \nolimits_{k=0}^{1}}%
%BeginExpansion
{\displaystyle\sum\nolimits_{k=0}^{1}}
%EndExpansion
\left\vert \triangle B_{0}\left\vert \psi_{B_{0}}^{\perp}\right\rangle
+(-1)^{k}\triangle B_{1}\left\vert \psi_{B_{1}}^{\perp}\right\rangle
\right\vert ^{2}}\nonumber\\
&  =\sqrt{2}\triangle A_{rms,2}\triangle B_{rms,2}, \label{ccsh}%
\end{align}
where $\left\langle CHSH\right\rangle _{L}=%
%TCIMACRO{\dsum \nolimits_{x,y=0}^{1}}%
%BeginExpansion
{\displaystyle\sum\nolimits_{x,y=0}^{1}}
%EndExpansion
(-1)^{xy}\left\langle A_{x}\right\rangle \left\langle B_{y}\right\rangle $,
and $\triangle X_{rms,n}=\sqrt{%
%TCIMACRO{\dsum \nolimits_{i=0}^{n-1}}%
%BeginExpansion
{\displaystyle\sum\nolimits_{i=0}^{n-1}}
%EndExpansion
(\triangle X_{i})^{2}}$. The equality of the first inequality in (\ref{ccsh})
holds if%

\begin{equation}
\left\vert \psi_{A_{x}}^{\perp}\right\rangle =\frac{\triangle B_{0}\left\vert
\psi_{B_{0}}^{\perp}\right\rangle +(-1)^{x}\triangle B_{1}\left\vert
\psi_{B_{1}}^{\perp}\right\rangle }{\left\vert \triangle B_{0}\left\vert
\psi_{B_{0}}^{\perp}\right\rangle +(-1)^{x}\triangle B_{1}\left\vert
\psi_{B_{1}}^{\perp}\right\rangle \right\vert }; \label{1}%
\end{equation}
and the equality of the second one holds if
\begin{equation}
\frac{\left\vert \triangle B_{0}\left\vert \psi_{B_{0}}^{\perp}\right\rangle
+\triangle B_{1}\left\vert \psi_{B_{1}}^{\perp}\right\rangle \right\vert
}{\triangle A_{0}}=\frac{\left\vert \triangle B_{0}\left\vert \psi_{B_{0}%
}^{\perp}\right\rangle -\triangle B_{1}\left\vert \psi_{B_{1}}^{\perp
}\right\rangle \right\vert }{\triangle A_{1}}. \label{2}%
\end{equation}
Consequently, we obtain the Tsirelson bound of the CHSH inequality%

\begin{align}
&  \left\langle CHSH\right\rangle \nonumber\\
&  \leq\left\langle CHSH\right\rangle _{L}+\sqrt{2}\triangle A_{rms,2}%
\triangle B_{rms,2},\nonumber\\
&  \leq2\sqrt{2}. \label{21}%
\end{align}
To derive the second inequality in (\ref{21}) as Tsirelson's bound of the CHSH
inequality, we begin by defining the function $f(\left\langle A_{0}%
\right\rangle ,\left\langle A_{1}\right\rangle ,\left\langle B_{0}%
\right\rangle ,\left\langle B_{1}\right\rangle )=%
%TCIMACRO{\dsum \nolimits_{x,y=0}^{1}}%
%BeginExpansion
{\displaystyle\sum\nolimits_{x,y=0}^{1}}
%EndExpansion
(-1)^{xy}\left\langle A_{x}\right\rangle \left\langle B_{y}\right\rangle
+\sqrt{2}\triangle A_{rms,2}\triangle B_{rms,2}$. Since $A_{x}^{2}=B_{y}%
^{2}=I$, we have $\triangle A_{rms,2}=\sqrt{%
%TCIMACRO{\dsum \nolimits_{x=0}^{1}}%
%BeginExpansion
{\displaystyle\sum\nolimits_{x=0}^{1}}
%EndExpansion
(1-\left\langle A_{x}\right\rangle ^{2})}$ and $\triangle B_{rms,2}=\sqrt{%
%TCIMACRO{\dsum \nolimits_{y=0}^{1}}%
%BeginExpansion
{\displaystyle\sum\nolimits_{y=0}^{1}}
%EndExpansion
(1-\left\langle B_{y}\right\rangle ^{2})}$. It can be verified that
$\frac{\partial f}{\partial\left\langle A_{0}\right\rangle }=\frac{\partial
f}{\partial\left\langle A_{1}\right\rangle }=\frac{\partial f}{\partial
\left\langle B_{0}\right\rangle }=\frac{\partial f}{\partial\left\langle
B_{1}\right\rangle }=0$ and $\frac{\partial^{2}f}{\partial\left\langle
A_{0}\right\rangle ^{2}},\frac{\partial^{2}f}{\partial\left\langle
A_{1}\right\rangle ^{2}},\frac{\partial^{2}f}{\partial\left\langle
B_{0}\right\rangle ^{2}},\frac{\partial^{2}f}{\partial\left\langle
B_{1}\right\rangle ^{2}}<0$ at $\left\langle A_{0}\right\rangle =\left\langle
A_{1}\right\rangle =\left\langle B_{0}\right\rangle =\left\langle
B_{1}\right\rangle =0$ ($\triangle A_{0}=\triangle A_{1}=\triangle
B_{0}=\triangle B_{1}=1$). Finally, we obtain Tsirelson's bound
\[
\underset{\left\vert \psi\right\rangle ,A_{x},B_{y}}{\max}f(\left\langle
A_{0}\right\rangle ,\left\langle A_{1}\right\rangle ,\left\langle
B_{0}\right\rangle ,\left\langle B_{1}\right\rangle )=2\sqrt{2},
\]
where $\widetilde{A}_{x}\left\vert \psi\right\rangle =A_{x}\left\vert
\psi\right\rangle =\triangle A_{x}\left\vert \psi_{A_{x}}^{\perp}\right\rangle
=\left\vert \psi_{A_{x}}^{\perp}\right\rangle $ and $\widetilde{B}%
_{y}\left\vert \psi\right\rangle =B_{y}\left\vert \psi\right\rangle =\triangle
B_{y}\left\vert \psi_{B_{y}}^{\perp}\right\rangle =\left\vert \psi_{B_{y}%
}^{\perp}\right\rangle $. In this case, Eq. (\ref{2}) indicates that either
$\left\langle \psi_{B_{0}}^{\perp}|\psi_{B_{1}}^{\perp}\right\rangle =0$ or
$\left\langle \psi_{B_{0}}^{\perp}|\psi_{B_{1}}^{\perp}\right\rangle =$ $\pm
i$. However, if $\left\langle \psi_{B_{0}}^{\perp}|\psi_{B_{1}}^{\perp
}\right\rangle =$ $\pm i$, we have $B_{1}\left\vert \psi\right\rangle
=\left\vert \psi_{B_{1}}^{\perp}\right\rangle =\pm i\left\vert \psi_{B_{0}%
}^{\perp}\right\rangle =$ $\pm iB_{0}\left\vert \psi\right\rangle $, which
indicates that the measurement outcome can be an imaginary number. Therefore,
we have $\left\langle \psi_{B_{0}}^{\perp}|\psi_{B_{1}}^{\perp}\right\rangle
=0$ and Eq. (\ref{1}) becomes
\begin{equation}
A_{x}\left\vert \psi\right\rangle =\frac{1}{\sqrt{2}}(B_{0}+(-1)^{x}%
B_{1})\left\vert \psi\right\rangle . \label{3}%
\end{equation}
In addition, Eq. (\ref{2}) can also be stated as $\left\langle \{B_{0}%
,B_{1}\}\right\rangle =0$, where $\{$ $,\}$ is the anti-commutator. By setting
$\{B_{0},B_{1}\}=0$, the equality $\left\langle \{B_{0},B_{1}\}\right\rangle
=0$ becomes state-independnet. Finally, to achieve Tsirelson's bound, we
assign the observables $B_{0}\rightarrow\sigma_{z}$, $B_{0}\rightarrow
\sigma_{x}$, $A_{0}\rightarrow\frac{1}{\sqrt{2}}(\sigma_{z}+\sigma_{x})$,
$A_{1}\rightarrow\frac{1}{\sqrt{2}}(\sigma_{z}-\sigma_{x})$; and the prepared
state $\left\vert \psi\right\rangle =\frac{1}{\sqrt{2}}(\left\vert
00\right\rangle +\left\vert 11\right\rangle )$.

A few remarks are in order. Firstly, the covariance of $A_{x}$ and $B_{y}$ is
defined as%

\[
\text{cov}\left\langle A_{x}B_{y}\right\rangle =\left\langle A_{x}%
B_{y}\right\rangle -\left\langle A_{x}\right\rangle \left\langle
B_{y}\right\rangle .
\]
Here we term $\left\langle \widetilde{CHSH}\right\rangle =%
%TCIMACRO{\dsum \nolimits_{x,y=0}^{1}}%
%BeginExpansion
{\displaystyle\sum\nolimits_{x,y=0}^{1}}
%EndExpansion
(-1)^{xy}$ cov$\left\langle A_{x}B_{y}\right\rangle $ CHSH covariance strength
\cite{var}. Eventually, (\ref{ccsh}) reveals the link among statisical
features such that the upper bound of the CHSH covariance strength is
proportional to the joint uncertainty. Secondly, given the condition
$\left\langle A_{0}\right\rangle =\left\langle A_{1}\right\rangle
=\left\langle B_{0}\right\rangle =\left\langle B_{1}\right\rangle =1$ and thus
$\triangle A_{0}=\triangle A_{1}=\triangle B_{0}=\triangle B_{1}=0$, Ineq.
(\ref{21}) becomes the CHSH inequality $\left\langle CHSH\right\rangle \leq2$.
In this case, $A_{x}\left\vert \psi\right\rangle =B_{y}\left\vert
\psi\right\rangle =\left\vert \psi\right\rangle $ $\forall x,y\in\{0,1\}$. A
possible trivial solution is that $\left\vert \psi\right\rangle $ is a product
tate $\left\vert \phi_{1}\right\rangle \otimes\left\vert \phi_{2}\right\rangle
$, and the operator $A_{0}=A_{1}$ $(B_{0}=B_{1})$ stabilizes $\left\vert
\phi_{1}\right\rangle \ $($\left\vert \phi_{2}\right\rangle $). Secondly,
using the Aharonov-Vaidman identity, the Pearson correlator of the observables
$A_{x}$ and $B_{y}$ can be written as $r(\widetilde{A}_{x}$, $\widetilde
{B}_{y})=\frac{\left\langle \widetilde{A}_{x}\widetilde{B}_{y}\right\rangle
}{\triangle A_{x}\triangle B_{y}}$ $=\left\langle \psi_{A_{x}}^{\perp}%
|\psi_{B_{y}}^{\perp}\right\rangle $. Denote Pearson's version of \ the
related CHSH value $\left\langle r\widetilde{CHSH}\right\rangle =%
%TCIMACRO{\dsum \nolimits_{x,y=0}^{1}}%
%BeginExpansion
{\displaystyle\sum\nolimits_{x,y=0}^{1}}
%EndExpansion
(-1)^{xy}r(\widetilde{A}_{x},\widetilde{B}_{y})$, and we have%

\begin{align}
&  \left\langle r\widetilde{CHSH}\right\rangle \nonumber\\
&  =%
%TCIMACRO{\dsum \nolimits_{x=0}^{1}}%
%BeginExpansion
{\displaystyle\sum\nolimits_{x=0}^{1}}
%EndExpansion
\left\langle \psi_{A_{x}}^{\perp}\right\vert (\left\vert \psi_{B_{0}}^{\perp
}\right\rangle +(-1)^{x}\left\vert \psi_{B_{1}}^{\perp}\right\rangle
)\nonumber\\
&  \leq\left\vert \left\vert \psi_{B_{0}}^{\perp}\right\rangle +\left\vert
\psi_{B_{1}}^{\perp}\right\rangle \right\vert +\left\vert \left\vert
\psi_{B_{0}}^{\perp}\right\rangle -\left\vert \psi_{B_{1}}^{\perp
}\right\rangle \right\vert \nonumber\\
&  =\sqrt{2+2\cos\lambda_{B}}+\sqrt{2-2\cos\lambda_{B}}\nonumber\\
&  \leq2\sqrt{2}, \label{4}%
\end{align}
where $\cos\lambda_{B}=\frac{1}{2}(\left\langle \psi_{B_{0}}^{\perp}%
|\psi_{B_{1}}^{\perp}\right\rangle +\left\langle \psi_{B_{1}}^{\perp}%
|\psi_{B_{0}}^{\perp}\right\rangle )$. The equality of the second inequality
in (\ref{4}) holds if $\cos\lambda_{B}=0$, which indicates that $\widetilde
{B}_{0}$ and $\widetilde{B}_{1}$ anti-commute. The result (\ref{4}) has also
been derived in \cite{var} but there is no statistical link between Bell
nonlocality and uncertainty. Thirdly, we can derive the result (\ref{ccsh}) in
an alternative way. Denote the Hermitian operators $\mathfrak{B}_{x}%
=B_{0}+(-1)^{x}B_{1}$, $x=0,1$. Note that $\left\langle \mathfrak{B}%
_{x}\right\rangle =\left\langle B_{0}\right\rangle +(-1)^{x}\left\langle
B_{1}\right\rangle $, and we have
\begin{align}
&  \left\langle CHSH\right\rangle -%
%TCIMACRO{\dsum \nolimits_{x,y=0}^{1}}%
%BeginExpansion
{\displaystyle\sum\nolimits_{x,y=0}^{1}}
%EndExpansion
(-1)^{xy}\left\langle A_{x}\right\rangle \left\langle B_{y}\right\rangle
\nonumber\\
&  =%
%TCIMACRO{\dsum \nolimits_{x=0}^{1}}%
%BeginExpansion
{\displaystyle\sum\nolimits_{x=0}^{1}}
%EndExpansion
\left\langle \widetilde{A_{x}}\widetilde{\mathfrak{B}_{x}}\right\rangle
\nonumber\\
&  =%
%TCIMACRO{\dsum \nolimits_{x=0}^{1}}%
%BeginExpansion
{\displaystyle\sum\nolimits_{x=0}^{1}}
%EndExpansion
\triangle A_{0}\triangle\mathfrak{B}_{0}\left\langle \psi_{A_{x}}^{\perp}%
|\psi_{\mathfrak{B}_{x}}^{\perp}\right\rangle \nonumber\\
&  \leq\sqrt{(\triangle A_{0})^{2}+(\triangle A_{1})^{2}}\sqrt{(\triangle
\mathfrak{B}_{x})\left\vert \left\langle \psi_{A_{x}}^{\perp}|\psi
_{\mathfrak{B}_{x}}^{\perp}\right\rangle \right\vert ^{2}}\nonumber\\
&  \leq\sqrt{(\triangle A_{0})^{2}+(\triangle A_{1})^{2}}\sqrt{(\triangle
\mathfrak{B}_{0})^{2}+(\triangle\mathfrak{B}_{1})^{2}}\nonumber\\
&  =\sqrt{2}\triangle A_{rms,2}\triangle B_{rms,2}. \label{444}%
\end{align}
Therein, the equality of the second inequality holds if $\left\vert
\left\langle \psi_{A_{0}}^{\perp}|\psi_{\mathfrak{B}_{0}}^{\perp}\right\rangle
\right\vert =\left\vert \left\langle \psi_{A_{1}}^{\perp}|\psi_{\mathfrak{B}%
_{1}}^{\perp}\right\rangle \right\vert =1$, and note that $(\triangle
\mathfrak{B}_{0})^{2}+(\triangle\mathfrak{B}_{1})^{2}=2(\triangle
B_{rms,2})^{2}$. Finally, (\ref{ccsh}) can be alternatively devised as the
form of the uncertainty relation
\begin{equation}
\sqrt{(\triangle A_{0})^{2}+(\triangle A_{1})^{2}}\sqrt{(\triangle B_{0}%
)^{2}+(\triangle B_{1})^{2}}\geq\frac{1}{\sqrt{2}}\left\langle \widetilde
{CHSH}\right\rangle , \label{55}%
\end{equation}
which states the joint uncertainty is lower bounded by the CHSH covariance
strength. Since a two-qubit entangled state is initially distributed between
Alice and Bob. There can never be a certain outcome for any local measurement.
Instead of evaluating the local uncertainty, they should evaluate the product
of RLSoVS as joint uncertainty, of which the lower bound is determined by CHSH
covariance strength. Note that, if CHSH covariance strength vanishes, it
allows that $\triangle A_{0}=\triangle A_{1}=\triangle B_{0}=\triangle
B_{1}=0$, which indicates that local outcomes each are constant.

As the generalized CHSH inequality, the chained Bell inequality reads
\cite{c1,c2}
\[
\left\langle \text{c-}Bell\right\rangle =\left\langle A_{0}B_{0}\right\rangle
-\left\langle A_{0}B_{n-1}\right\rangle +%
%TCIMACRO{\dsum \nolimits_{k=1}^{n-1}}%
%BeginExpansion
{\displaystyle\sum\nolimits_{k=1}^{n-1}}
%EndExpansion
(\left\langle A_{k}B_{k-1}\right\rangle +\left\langle A_{k}B_{k}\right\rangle
).
\]
This value is obtained by allowing for more than two measurement settings for
Alice and Bob. Similarly, denote the related chained Bell value $\left\langle
\widetilde{\text{c-}Bell}\right\rangle =\left\langle \widetilde{A_{0}%
}\widetilde{B_{0}}\right\rangle -\left\langle \widetilde{A_{0}}\widetilde
{B_{n-1}}\right\rangle +%
%TCIMACRO{\dsum \nolimits_{k=1}^{n-1}}%
%BeginExpansion
{\displaystyle\sum\nolimits_{k=1}^{n-1}}
%EndExpansion
\left\langle \widetilde{A_{k}}B\widetilde{_{k-1}}\right\rangle +\left\langle
\widetilde{A_{k}}\widetilde{B_{k}}\right\rangle $, and using the similar
approach of deriving (\ref{21}), we have%

\[
\left\langle \widetilde{\text{c-}Bell}\right\rangle \leq\sqrt{2}\triangle
A_{rms,n}\sqrt{(\triangle B_{rms,n})^{2}+%
%TCIMACRO{\dsum \nolimits_{j=0}^{n-1}}%
%BeginExpansion
{\displaystyle\sum\nolimits_{j=0}^{n-1}}
%EndExpansion
\triangle B_{j}\triangle B_{j+1}\cos\lambda_{j}},
\]
or equivalently,
\begin{align}
\left\langle \text{c-}Bell\right\rangle -\left\langle \text{c-}%
Bell\right\rangle _{L}  &  \leq\sqrt{2}\triangle A_{rms,n}\sqrt{(\triangle
B_{rms,n})^{2}+%
%TCIMACRO{\dsum \nolimits_{j=0}^{n-1}}%
%BeginExpansion
{\displaystyle\sum\nolimits_{j=0}^{n-1}}
%EndExpansion
\triangle B_{j}\triangle B_{j+1}\cos\lambda_{j}}\nonumber\\
&  \leq\sqrt{2}\triangle A_{rms,n}\sqrt{(\triangle B_{rms,n})^{2}+%
%TCIMACRO{\dsum \nolimits_{j=0}^{n-1}}%
%BeginExpansion
{\displaystyle\sum\nolimits_{j=0}^{n-1}}
%EndExpansion
\triangle B_{j}\triangle B_{j+1}}, \label{22}%
\end{align}
where $\left\langle \text{c-}Bell\right\rangle _{L}=$ $\left\langle
A_{0}\right\rangle \left\langle B_{0}\right\rangle -\left\langle
A_{0}\right\rangle \left\langle B_{n-1}\right\rangle +%
%TCIMACRO{\dsum \nolimits_{k=1}^{n-1}}%
%BeginExpansion
{\displaystyle\sum\nolimits_{k=1}^{n-1}}
%EndExpansion
\left\langle A_{k}\right\rangle \left\langle B_{k-1}\right\rangle
+\left\langle A_{k}\right\rangle \left\langle B_{k}\right\rangle $,
$\cos\lambda_{j}=\frac{1}{2}(\left\langle \psi_{B_{j}}^{\perp}|\psi_{B_{j+1}%
}^{\perp}\right\rangle +\left\langle \psi_{B_{j+1}}^{\perp}|\psi_{B_{j}%
}^{\perp}\right\rangle )$, $0\leq j\leq n-2$, $\triangle B_{n}=\triangle
B_{0},\cos\lambda_{n-1}=\frac{-1}{2}(\left\langle \psi_{B_{0}}^{\perp}%
|\psi_{B_{n-1}}^{\perp}\right\rangle +\left\langle \psi_{B_{n-1}}^{\perp}%
|\psi_{B_{0}}^{\perp}\right\rangle )$. A few remarks are in order. It is easy
to verify that $\left\langle \text{c-}Bell\right\rangle _{L}\leq2n-2$, where
the equality holds if $\left\langle A_{x}\right\rangle =\left\langle
B_{y}\right\rangle =1$ and hence $\triangle A_{x}=\triangle B_{y}=0$.
Consequently, it leads to the chained Bell inequality
\[
\left\langle \text{c-}Bell\right\rangle \overset{LHV}{\leq}2n-2=\max
\left\langle \text{c-}Bell\right\rangle _{L}%
\]
Secondly, if the specific Bell state is distributed between Alice a Bob, the
statistical results $\left\langle A_{x}\right\rangle =\left\langle
B_{y}\right\rangle =0$ and hence $\triangle A_{x}=\triangle B_{y}=1$ leads to
$\left\langle \text{chain-}Bell\right\rangle \leq\sqrt{2n}\sqrt{n+%
%TCIMACRO{\dsum \nolimits_{j=0}^{n-1}}%
%BeginExpansion
{\displaystyle\sum\nolimits_{j=0}^{n-1}}
%EndExpansion
\cos\lambda_{j}}.$ In addition, given $\cos\lambda_{0}=\cdots=\cos
\lambda_{n-1}=\cos\lambda$, we have $\left\langle \text{c-}Bell\right\rangle
=2n\cos\frac{\lambda}{2}$, and the Tsirelson bound can be obtained by setting
$\lambda=\frac{\pi}{n}$. That is, in (\ref{22}), the Tsirelson bound of the
chained Bell inequality cannot be determined simply by the statistics of the
local observables like (\ref{ccsh}) and (\ref{21}). Eventually, the maximum
violation in the post-quantum region can be achieved by setting $\left\langle
A_{k}B_{k-1}\right\rangle =\left\langle A_{k}B_{k}\right\rangle =\left\langle
A_{0}B_{0}\right\rangle =-\left\langle A_{0}B_{n-1}\right\rangle =1$ and
$\cos\lambda=1$, which leads to the trivial inequality $\left\langle
\text{c-}Bell\right\rangle \leq2n$. However, the maximal violation of the CHSH
inequality in the post-quantum region is unattainable in (\ref{ccsh}),
(\ref{21}), and (\ref{444}).

\textit{N spin-s particles case }Denote recursive $n$-qudit Bell operators of
Mermin-Klyshko inequality by \cite{MK, cabello}
\[
B_{n}=B_{k}B_{n-k}^{+}+B_{k}^{\prime}B_{n-k}^{-},
\]
and
\[
B_{n}^{\prime}=B_{n-k}^{\prime}B_{k}^{+}-B_{n-k}B_{k}^{-},
\]
where $B_{n-k}^{\pm}=B_{n-k}\pm$ $B_{n-k}^{\prime}$ and lettinng one-qudit
opertaors $B_{1}=A_{1}$, $B_{1}^{\prime}=A_{1}^{\prime}$. Using a similar
approach in (\ref{ccsh}) with replacing $A_{0}$, $A_{1}$, $B_{0}$ and $B_{1}$
by $B_{k}$, $B_{k}^{\prime}$, $B_{n-k}$ and $B_{n-k}^{\prime}$, respectively,
we have%

\[
\left\langle B_{n}\right\rangle \leq F(\{\left\langle B_{m}\right\rangle
,\left\langle B_{m}^{\prime}\right\rangle \},m=k,n-k),
\]
where the function $F(\{\left\langle B_{m}\right\rangle ,\left\langle
B_{m}^{\prime}\right\rangle \},m=k,n-k)=\left\langle B_{n}\right\rangle
_{L}+\sqrt{2}B_{rms}^{(k)}B_{rms}^{(n-k)}$, $\left\langle B_{n}\right\rangle
_{L}=\left\langle B_{k}\right\rangle (\left\langle B_{n-k}\right\rangle
+\left\langle B_{n-k}^{\prime}\right\rangle )+\left\langle B_{k}^{\prime
}\right\rangle (\left\langle B_{n-k}\right\rangle -\left\langle B_{n-k}%
^{\prime}\right\rangle )$ and $B_{rms}^{(m)}=\sqrt{(\triangle B_{m}%
)^{2}+(\triangle B_{m}^{\prime})^{2}}$. \ To find the Tsirelson bound of
$\left\langle B_{n}\right\rangle $, it is obvious that $\frac{\partial
F}{\partial\left\langle B_{k}\right\rangle }=\frac{\partial F}{\partial
\left\langle B_{k}^{\prime}\right\rangle }=\frac{\partial F}{\partial
\left\langle B_{n-k}\right\rangle }=\frac{\partial F}{\partial\left\langle
B_{n-k}^{\prime}\right\rangle }=0$ and $\frac{\partial^{2}F}{\partial
\left\langle B_{k}\right\rangle ^{2}},\frac{\partial^{2}F}{\partial
\left\langle B_{k}^{\prime}\right\rangle ^{2}},\frac{\partial^{2}F}%
{\partial\left\langle B_{n-k}\right\rangle ^{2}},\frac{\partial^{2}F}%
{\partial\left\langle B_{n-k}^{\prime}\right\rangle }<0$ at $\left\langle
B_{k}\right\rangle =\left\langle B_{k}^{\prime}\right\rangle =\left\langle
B_{n-k}\right\rangle =\left\langle B_{n-k}^{\prime}\right\rangle =0$. In
addition, we can utilize the following lemma.

\textit{Lemma 1} $B_{m}^{2}=B_{m}^{\prime2}\geq0$ and $\max tr(B_{m}%
^{2})=2^{3(m-1)}$.

Detailed proof can be found in \cite{bellsquare}. Consequently, we have
\begin{align}
&  \left\langle B_{n}\right\rangle \nonumber\\
&  \leq\sqrt{2}%
%TCIMACRO{\dsum \nolimits_{m=k,n-k}}%
%BeginExpansion
{\displaystyle\sum\nolimits_{m=k,n-k}}
%EndExpansion
\sqrt{tr(B_{m}^{2})+tr(B_{m}^{\prime2})}\nonumber\\
&  \leq\sqrt{2}%
%TCIMACRO{\dsum \nolimits_{m=k,n-k}}%
%BeginExpansion
{\displaystyle\sum\nolimits_{m=k,n-k}}
%EndExpansion
\sqrt{2\times2^{3(m-1)}}\nonumber\\
&  =2^{\frac{3(n-1)}{2}}. \label{BB}%
\end{align}
The first inequality in (\ref{BB}) is due to the fact $\left\langle B_{m}%
^{2}\right\rangle \leq tr(B_{m}^{2})$, and the equality holds if there is only
one non-zero eigenvalue $2^{3(m-1)}$. In the end, the statistical result
$\triangle B_{k}=\triangle B_{k}^{\prime}=\triangle B_{n-k}=\triangle
B_{n-k}^{\prime}=0$ leads to Mermin-Klyshko inequalities \cite{cabello}%

\begin{equation}
\left\langle B_{n}\right\rangle \overset{LHV}{\leq}\max\left\langle
B_{n}\right\rangle _{L}=2^{n-1}.\nonumber
\end{equation}

In conclusion, we open the way for establishing the statistical link between
uncertainty relations and Bell nonlocality. Therein, the Aharonov-Vaidman
identity makes it possible to express the upper bounds of the experimental
Bell values in terms of statistical characters such as expectation values and
variances of local observables. As a result, the joint uncertainty determines
the upper bound of Bell inequalities, and the sum-of-correlation determines
the lower bound of the joint uncertainty.

There are some interesting avenues for further study. We expect that the
Aharonov-Vaidman identity can be used to establish the link between the
uncertainty relations and other nonlocal properties such as the steering
effect and quantum entanglement \cite{steer,steer2}. It would also be
worthwhile to explore this statistical connection in quantum networks
\cite{net}. Lastly, it is unknown whether there exists an identity that
reveals the statistical characteristics of non-Hermitian observables on a
state vector. If such an identity exists, it would be intriguing to seek the
link between the uncertainty relations and nonlocal effects of non-Hermitian
observables \cite{non,nnon}.

The work is financially supported by National Science and Technology Council
(NSTC) with Grant No. NSTC 113-2112-M-033 -006.


\begin{thebibliography}{99}
\bibitem {Un}However, if $\left\vert \psi\right\rangle $ be a single-qubit
state and the obervables $A$ and $B$ on the same qubit, $\left\langle
\widetilde{A}\widetilde{B}\right\rangle _{\psi}$ and hence $\left\langle
\psi_{A}^{\perp}|\psi_{B}^{\perp}\right\rangle $ can be complex number. For
example, let $A$ and $B$ be $\sigma_{x}$ and $\sigma_{z}$, respectively. In
fact, since a qubit is a two-level quantum system, it is easy to verified that
$\left\vert \psi_{A}^{\perp}\right\rangle =e^{i\phi_{AB}}\left\vert \psi
_{B}^{\perp}\right\rangle $ for some overall phase $\phi_{AB}$.
\end{thebibliography}
\end{document}